\documentclass[aps,pra,showpacs,superscriptaddress]{revtex4}%
\usepackage{amsmath,amssymb}

\begin{document}

\title{Greenberger-Horne-Zeilinger argument of nonlocality without inequalities for mixed states.}

\author{GianCarlo \surname{Ghirardi}}\email{ghirardi@ts.infn.it}%
\affiliation{Department of Theoretical Physics, University of
  Trieste, Italy}%
\affiliation{Istituto Nazionale di Fisica Nucleare, Sezione di Trieste,
  Italy}%
\affiliation{International Centre for Theoretical Physics ``Abdus Salam,''
  Trieste, Italy}%

\author{Luca \surname{Marinatto}}\email{marinatto@ts.infn.it}%
\affiliation{Department of Theoretical Physics, University of
  Trieste, Italy}%
\affiliation{Istituto Nazionale di Fisica Nucleare, Sezione di Trieste,
  Italy}%

\date{\today}

\begin{abstract}
We generalize the Greenberger-Horne-Zeilinger nonlocality without inequalities argument
 to cover the case of arbitrary mixed statistical operators associated to three-qubits
 quantum systems.
More precisely, we determine the radius of a ball (in the trace distance topology)
 surrounding the pure GHZ state and containing arbitrary mixed statistical operators
 which cannot be described by any local and realistic hidden variable model and which
 are, as a consequence, noncompletely separable.
As a practical application, we focus on certain one-parameter classes of mixed states which
 are commonly considered in the experimental realization of the original GHZ argument
 and which result from imperfect preparations of the pure GHZ state.
In these cases we determine for which values of the parameter measuring the noise
 a nonlocality argument can still be exhibited, despite the mixedness of the considered
 states. Moreover, the effect of the imperfect nature of measurement processes is
 discussed.
\end{abstract}


\maketitle

\section{Introduction}

As is well known, the standard way to put into evidence the nonlocal features associated to pure entangled
 states is to resort to Bell-like inequalities~\cite{bell,chsh} involving linear combinations of correlation
 functions and to show that they are experimentally violated.
However, this approach, in general, may turn out to be difficult both from the theoretical (identifying the
 appropriate class of Bell's inequalities) and from the experimental (collecting the data referring to the
 correlated outcomes) point of view.
A significant improvement concerning this issue has been achieved by devising clever logical arguments pointing
 up a conflict between any conceivable local realistic hidden variable model and the quantum
 mechanical predictions without resorting to inequalities. The most relevant examples of this line of thought are
 the celebrated Greenberger-Horne-Zeilinger~\cite{ghz} (GHZ, in what follows) and Hardy's~\cite{hardy}
 proofs which hold for an appropriate pure state of a tripartite spin-$1/2$ composite system (the GHZ state,
 in what follows) and for almost all bipartite pure entangled states in $\mathbb{C}^{2}\otimes\mathbb{C}^{2}$,
 respectively.

The situation turns out to be much more complicated when consideration is given to mixed states. In fact, on one
 side, in this case the occurrence of entanglement does not guarantee, by itself, that there exists a Bell's
 inequality which is violated and, as a consequence that no local realistic model can reproduce the quantum
 predictions; actually, there exist mixed entangled states admitting a description in terms of local
 variables~\cite{werner}.
Accordingly, to tackle the locality problem in the case of mixed entangled states, new approaches might be
 useful.
In this spirit we stress that quite recently, Hardy's argument~\cite{hardy} has been generalized to cover
 firstly~\cite{gm1} the case of multipartite pure states and, subsequently, the case of a large class of
 bipartite mixed states~\cite{gm2}, both proofs holding for systems with Hilbert spaces of arbitrary dimensionality.
The main purpose of this paper is to generalize, by following the same line of reasoning and techniques
 involving simple set theoretic arguments similar to those of Refs.~\cite{gm1,gm2}, the original GHZ
 argument~\cite{ghz} to mixed states associated to tripartite spin-$1/2$ composite systems.
More precisely, for any statistical operator $\varrho$ describing a system of this type we will derive a
 constraint for its trace distance from the projection operator onto the GHZ state, whose violation implies that
 no local realistic model exists reproducing the quantum prediction implied by the state $\varrho$.
Thus, the measurement outcomes on such entangled states are genuinely nonlocally correlated and one is
 led to the identification of a precise spherical neighborhood (in the topology induced by the trace distance)
 of the pure GHZ state such that all statistical operators belonging to it cannot be completely separable, i.e.
 they cannot be expressed as a convex sum of one-particle product states. Our result is relevant for  quantum
information processing because the identification of a class of mixed
 entangled states which cannot be mimicked by any local realistic model necessarily implies
 that no classical and local communication protocol exists which can outperform the quantum
 protocols exploiting the specific quantum features of such states~\cite{brassard,toner}.

There is another aspect for which our proof might turn out to be relevant. From a practical point of view any
 test of nonlocality along the GHZ line of thought~\cite{bouwmeester,pan} requires the verification of perfect
 (anti)correlations between the outcomes of appropriate measurements, involving three spacelike separated
 particles.
So far, the most widely used and most reliable source of multipartite entanglement is represented
 by polarization-entangled photons produced in parametric down-conversion experiments~\cite{kwiat}.
Unfortunately, in these experiments perfect (anti)correlations can never be really observed due to
 practical limitations deriving both from imperfect preparations of a pure GHZ state
 and from the unavoidably imperfect orientations  of the single-particle detectors.
Accordingly, the arguments of this paper also lead to identify  under which circumstances one can
 still put into evidence nonlocal effects despite the presence of the above-mentioned practical limitations.
For the sake of definiteness, we will explicitly consider  two different classes of one-parameter
 mixed states (describing mixtures of a pure GHZ state with two different kind of noise, a white and a
 colored one) and we will determine the exact range of values of the parameter measuring the amount of noise
 for which  the resulting mixed states exhibit genuine and detectable nonlocal effects. An analogous analysis
 will be developed with reference to the issue of imperfect measurements which will lead us to determine which
 degree of imperfection will not spoil completely the possibility of revealing nonlocal effects.


\section{Hidden variable models}

The {\em completeness assumption} of the Copenhagen version of quantum mechanics
 postulates that all what one can know about a quantum physical system
 is determined by the quantum state (represented by a normalized vector of a Hilbert space).
By denying such a postulate, one opens the possibility that the statistical
 distributions connected with the outcomes of quantum measurements, which have been confirmed
 beyond every reasonable doubt by innumerable experiments, can be thought of as arising
 because the quantum states do not represent the maximal possible knowledge one can have
 about a quantum system.
If one takes this position the results of any conceivable individual quantum measurement could be
 thought as predetermined by the specification of additional hidden variables
 (hidden because if one could really prepare a quantum state with a particular value of
 such variables, quantum mechanics would turn out to be experimentally inadequate)
 in a such a way that an average over them reproduces the quantum mechanical probabilities.

The hypothetical theory completing quantum mechanics for a n-partite system is generally
 referred to as a stochastic hidden variable model, and it consists of:
 (i) a set $\Lambda$ whose elements $\lambda$ are called hidden variables; (ii) a normalized
 probability distribution $\rho$ defined on $\Lambda$; (iii) a set of
 probability distributions $P_{\lambda}(A_{i}\!=\!a, B_{j}\!=\!b,\dots, Z_{k}\!=\!z)$
 for the outcomes of single and joint measurements of any conceivable
 set of observables $\left\{A_{i},B_{j},\dots,Z_{k} \right\}$
 where each index of the set $\left\{ i,j,\dots,k\right\}$ refers to a single particle or to
 a group of all the $n$ particles, such that
\begin{equation}
\label{eq0.1} P_{\varrho}(A_{i}=a,B_{j}=b,\dots,Z_{k}=z) = \int_{\Lambda}\,d\lambda\,
   \rho(\lambda) P_{\lambda}(A_{i}=a,B_{j}=b,\dots,Z_{k}=z).
\end{equation}
The quantity at the left hand side is the probability distribution
 which quantum mechanics attaches to the outcomes $\left\{ a,b, \dots, z\right\}$ of the
 considered measurements when the system is described by the (pure or mixed) state $\varrho$.
A deterministic hidden variable model, also known as realistic model, constitutes a
 particular instance of a stochastic one where all probabilities $P_{\lambda}$ take only
 the values $0$ or $1$.
In this last mentioned situation, the experimentally verified predictions of quantum mechanics
 arise as averages over the distribution of the hidden variables (which are commonly
 referred to as dispersion-free states~\cite{bellRMP}), and quantum mechanics turns out to be
 the theory emerging from a more fundamental deterministic theory (as happens with statistical
 classical mechanics with respect to classical mechanics).

A hidden variable model is called local~\cite{bell2} if the following factorizability condition
 holds for any conceivable joint probability distribution $P_{\lambda}(A_{i}=a,B_{j}=b, \ldots, Z_{k}=z)$
 and for any value of the hidden variable $\lambda\in \Lambda$,
\begin{equation}
 \label{eq0.2}
 P_{\lambda}(A_{i}=a,B_{j}=b,\ldots,Z_{k}=z)= P_{\lambda}(A_{i}=a)P_{\lambda}(B_{j}=b)\ldots
 P_{\lambda}(z_{k}=z)
 \end{equation}
in all cases in which the measurement processes for the observables $A_{i},B_{j},\ldots, Z_{k}$
 occur at spacelike separated locations. The locality condition imposes that no causal influence
 can exist between spacelike separated events.
Since deterministic and stochastic hidden variable models are totally equivalent when the locality
 condition is imposed~\cite{fine}, and since in what follows we will focus only on realistic models,
 all the probabilities $P_{\lambda}$ will consequently be assumed to take the values $0$ and $1$ only.


\section{Generalized GHZ proof of nonlocality}

To start with, let us define the three-qubit GHZ state we will use in this paper as
\begin{equation}
\label{eq1.1}
 \vert GHZ\rangle= \frac{1}{\sqrt{2}} (\vert 0 \rangle \otimes |0\rangle \otimes|0\rangle +
 \vert 1 \rangle\otimes |1\rangle \otimes |1\rangle),
\end{equation}
where we have denoted as $\vert 0\rangle$ and $\vert 1\rangle$ the eigenstates of the Pauli matrix
 $\sigma_{z}$ associated to the eigenvalues $+1$ and $-1$, respectively.
Now consider the set of four observables
 $\sigma_{1x}\otimes \sigma_{2x} \otimes \sigma_{3x}$,
 $\sigma_{1x}\otimes \sigma_{2y} \otimes \sigma_{3y}$,
 $\sigma_{1y}\otimes \sigma_{2x} \otimes \sigma_{3y}$ and
 $\sigma_{1y}\otimes \sigma_{2y} \otimes \sigma_{3x}$, tensor products
 of the Pauli matrices $\sigma_{i}$ ($i=x,y,z$).
Given the state of Eq.~(\ref{eq1.1}), the following joint probability distributions easily follow:
\begin{eqnarray}
\label{eq1.21}
 P_{GHZ}(\sigma_{1x}\otimes \sigma_{2x} \otimes \sigma_{3x}=+1) & = & 1, \\
 \label{eq1.22}
P_{GHZ}(\sigma_{1x}\otimes \sigma_{2y} \otimes \sigma_{3y}=-1) & = & 1,\\
\label{eq1.23}
P_{GHZ}(\sigma_{1y}\otimes \sigma_{2x} \otimes \sigma_{3y}=-1) & = & 1, \\
\label{eq1.24}
 P_{GHZ}(\sigma_{1y}\otimes \sigma_{2y} \otimes \sigma_{3x}=-1) & = & 1.
\end{eqnarray}
In the original GHZ argument~\cite{ghz} it is shown that the the existence of a local and realistic
 hidden variable model for the state of Eq.~(\ref{eq1.1}) cannot reproduce simultaneously
 all the predictions of Eqs.~(\ref{eq1.21})-(\ref{eq1.24}). Stated equivalently, the perfect
 (anti)correlations between the outcomes of certain spin-observables implied by the probability
 distributions (\ref{eq1.21})-(\ref{eq1.24}) are said to be locally inexplicable.
Now, consider an arbitrary tripartite statistical operator $\varrho$ acting in
 the factorized Hilbert space $\mathbb{C}^{2}\otimes\mathbb{C}^{2}\otimes\mathbb{C}^{2}$
 and denote with the non-negative parameter $\varepsilon$ its trace distance
 $D(\varrho, \vert GHZ \rangle \langle GHZ \vert)$ from the pure state
 $\vert GHZ \rangle \langle GHZ \vert$, that is
\begin{equation}
 \label{eq1.30}
 D(\varrho, \vert GHZ \rangle \langle GHZ \vert) \equiv \frac{1}{2} \textrm{Tr} \vert\:(
  \varrho -\vert GHZ \rangle \langle GHZ \vert )\:\vert = \varepsilon,
\end{equation}
where $\vert A \vert \equiv \sqrt{A^{\dagger}A}$ is the positive square root of
 the operator $A^{\dagger}A$.
The trace distance $D(\delta,\bar{\delta})$ between two arbitrary statistical operators
 $\delta$ and $\bar{\delta}$ represents a good measure to quantify the closeness of such states.
Moreover, a well known property of the trace distance implies that if two states are close in
 the trace distance, then they will give rise to probability distributions for the outcomes
 of arbitrary observables which are close together too.
In fact, the following relation holds
\begin{equation}
\label{eq1.3} \vert \textrm{Tr}[P\delta] - \textrm{Tr}[P\bar{\delta}]\vert \leq D(\delta,\bar{\delta})
\end{equation}
for any projection operator $P$ (projecting onto a monodimensional or onto a multidimensional
 manifold), the expression $\textrm{Tr}[P\delta]$ and $\textrm{Tr}[P\bar{\delta}]$
 representing the probability for the occurrence of the measurement outcome associated to $P$
 when the system is in the state $\delta$ and $\bar{\delta}$, respectively.
Due to this property, we may confidently expect that it must exist a neighborhood (in the trace distance
 topology) surrounding the pure GHZ state, which consists of mixed states: (i) satisfying probability
 distributions which are close to those of Eqs.~(\ref{eq1.21})-(\ref{eq1.24}) and, as a
 consequence, (ii) which exhibit the same nonlocal effects of the GHZ state itself.
Thus, by making use of Eq.~(\ref{eq1.3}), we may
 calculate how the probability distributions of Eqs.~(\ref{eq1.21})-(\ref{eq1.24}) are modified
 when the state of the system is the mixed state $\varrho$ of Eq.~(\ref{eq1.30}), obtaining
\begin{eqnarray}
\label{eq1.41}
 P_{\varrho}(\sigma_{1x}\otimes \sigma_{2x} \otimes \sigma_{3x}=+1) & \geq & 1 -\varepsilon, \\
 \label{eq1.42}
P_{\varrho}(\sigma_{1x}\otimes \sigma_{2y} \otimes \sigma_{3y}=-1) & \geq & 1 -\varepsilon, \\
\label{eq1.43}
P_{\varrho}(\sigma_{1y}\otimes \sigma_{2x} \otimes \sigma_{3y}=-1) & \geq & 1 -\varepsilon, \\
\label{eq1.44}
 P_{\varrho}(\sigma_{1y}\otimes \sigma_{2y} \otimes \sigma_{3x}=-1) & \geq & 1 -\varepsilon.
\end{eqnarray}
Let us now suppose that a local and deterministic hidden variable model, as defined in
 the previous section and reproducing the quantum mechanical probabilities $P_{\varrho}$
 of Eqs.~(\ref{eq1.41})-(\ref{eq1.44}), exists for $\varrho$.
This implies that, for example, Eq.~(\ref{eq1.41}) becomes
\begin{eqnarray}
\label{eq1.5}
 P_{\varrho}(\sigma_{1x}\otimes \sigma_{2x} \otimes \sigma_{3x}=+1) & = &
 \sum_{rst=+1}P_{\varrho}(\sigma_{1x}=r,\sigma_{2x}=s,\sigma_{3x}=t) \\
 \label{eq1.51}
& = &\sum_{rst=+1}\int_{\Lambda}d \lambda \rho(\lambda) P_{\lambda}(\sigma_{1x}=r, \sigma_{2x}=s,
 \sigma_{3x}=t)\\
 \label{eq1.52}
 & = &\int_{\Lambda}d \lambda \rho(\lambda)\sum_{rst=+1} P_{\lambda}(\sigma_{1x}=r) P_{\lambda}(\sigma_{2x}=s)
 P_{\lambda}( \sigma_{3x}=t) \geq 1 -\varepsilon,
\end{eqnarray}
where the indices $r,s,t$ may assume the values $+1$ and $-1$ only and they are constrained by the
 relation $rst=+1$ since we are interested in the outcome $+1$ for the observable
 $\sigma_{1x}\otimes \sigma_{2x} \otimes \sigma_{3x}$. Moreover, the third equality comes from the
 locality condition of Eq.~(\ref{eq0.2}).
Since we are dealing with a deterministic model where all probabilities $P_{\lambda}$ are
 equal to $0$ or $1$, and since for any $\lambda$ at most one term in the sum
 $\sum_{rst=+1}P_{\lambda}(\sigma_{1x}=r) P_{\lambda}(\sigma_{2x}=s)P_{\lambda}( \sigma_{3x}=t)$
 can be different from zero, the sum itself can take only the values $0$ or $1$.
Thus, given the set $\Lambda$ of the hidden variables, we define the following subset $E$
 of $\Lambda$ as follows:
\begin{equation}
\label{eq1.6}
 E =  \left\{ \lambda\in \Lambda \:\vert\:\sum_{rst=+1}P_{\lambda}(\sigma_{1x}=r) P_{\lambda}(\sigma_{2x}=s)
 P_{\lambda}( \sigma_{3x}=t)=1
  \right\}.
\end{equation}
As a consequence, Eq.~(\ref{eq1.52}) takes the simpler form:
\begin{equation}
\label{eq1.7}
 \int_{\Lambda}d \lambda \rho(\lambda)\sum_{rst=+1} P_{\lambda}(\sigma_{1x}=r)P_{\lambda}(\sigma_{2x}=s)
 P_{\lambda}( \sigma_{3x}=t)=\int_{E}d \lambda \rho(\lambda) = \mu[E] \geq 1-\varepsilon,
\end{equation}
where we have denoted as $\mu[E]$ the measure of the subset $E$ with respect to the probability
 distribution $\rho(\lambda)$.
In an analogous manner, by defining the following subsets $F,G$, and $H$ of $\Lambda$:
\begin{eqnarray}
\label{eq1.61}
 F &=&  \left\{ \lambda\in \Lambda \:\vert\:\sum_{rst=-1}P_{\lambda}(\sigma_{1x}=r) P_{\lambda}(\sigma_{2y}=s)
 P_{\lambda}( \sigma_{3y}=t)=1  \right\},\\
\label{eq1.62}
 G &=&  \left\{ \lambda\in \Lambda \:\vert\:\sum_{rst=-1}P_{\lambda}(\sigma_{1y}=r) P_{\lambda}(\sigma_{2x}=s)
 P_{\lambda}( \sigma_{3y}=t)=1  \right\},\\
 \label{eq1.63}
 H &=&  \left\{ \lambda\in \Lambda \:\vert\:\sum_{rst=-1}P_{\lambda}(\sigma_{1y}=r) P_{\lambda}(\sigma_{2y}=s)
 P_{\lambda}( \sigma_{3x}=t)=1  \right\},
\end{eqnarray}
we may rewrite Eqs.~(\ref{eq1.41})-(\ref{eq1.44}) in a simpler form in
 terms of the measure of the subsets $E,F,G$, and $H$ of $\Lambda$:
\begin{eqnarray}
\label{eq1.801}
\mu [ E ] & \geq &1-\varepsilon, \\
 \label{eq1.802}
 \mu [ F ] & \geq &1-\varepsilon, \\
 \label{eq1.803}
 \mu [ G ] & \geq &1-\varepsilon, \\
 \label{eq1.804}
 \mu [ H ] & \geq &1-\varepsilon.
\end{eqnarray}
Now, consider a fixed value of $\lambda$ belonging, e.g., to the subset $E$: given this maximal
 specification of the state of the system and by the very definition of $E$, the relation
 $\sum_{rst=+1}P_{\lambda}(\sigma_{1x}=r) P_{\lambda}(\sigma_{2x}=s)P_{\lambda}( \sigma_{3x}=t)=1$ holds.
This may happen only when (exactly) one addendum of the sum is equal to one; this in turn implies
 that, for any specific value $\lambda \in E$, only one of the following four set of equalities holds:
 \begin{eqnarray}
 \label{eq1.8001}
 P_{\lambda}(\sigma_{1x}=+1) & = & P_{\lambda}(\sigma_{2x}=+1)= P_{\lambda}( \sigma_{3x}=+1)=1,\\
 \label{eq1.8002}
P_{\lambda}(\sigma_{1x}=+1) & = & P_{\lambda}(\sigma_{2x}=-1)= P_{\lambda}( \sigma_{3x}=-1)=1,\\
 \label{eq1.8003}
P_{\lambda}(\sigma_{1x}=-1) & = & P_{\lambda}(\sigma_{2x}=+1)= P_{\lambda}( \sigma_{3x}=-1)=1,\\
 \label{eq1.8004}
P_{\lambda}(\sigma_{1x}=-1) & = & P_{\lambda}(\sigma_{2x}=-1)= P_{\lambda}( \sigma_{3x}=+1)=1.
 \end{eqnarray}
In a deterministic theory, like the one we are dealing with, since all (single-particle) probabilities
 are equal to $0$ or $1$, the measurement outcomes of the associated observables are predetermined for
 any known value of $\lambda$. We denote such values as $v_{\lambda}(\sigma_{ij})$,
 where $i=1,2,3$ is the particle index, while the $j=x,y,z$ refers to the spin-component.
For example, Eq.~(\ref{eq1.8002}) implies that $v_{\lambda}(\sigma_{1x})=+1$ while
 $v_{\lambda}(\sigma_{2x})=v_{\lambda}(\sigma_{3x})=-1$ and, moreover,
 $v_{\lambda}(\sigma_{1x})v_{\lambda}(\sigma_{2x})v_{\lambda}(\sigma_{3x})=+1$ .
Similarly, as a consequence of Eqs.~{(\ref{eq1.8001})-(\ref{eq1.8004}),
 for any $\lambda \in E$ the following constraint has to be satisfied:
\begin{equation}
\label{eq1.9}
 v_{\lambda}(\sigma_{1x})v_{\lambda}(\sigma_{2x})v_{\lambda}(\sigma_{3x})=+1.
\end{equation}
By arguing in the same way for all the remaining subsets $F,G$, and $H$ of $\Lambda$, we obtain
\begin{eqnarray}
\label{eq5.61}
 v_{\lambda}(\sigma_{1x})v_{\lambda}(\sigma_{2x})v_{\lambda}(\sigma_{3x}) & = & +1\hspace{0.7cm}\lambda \in E,\\
\label{eq5.62}
 v_{\lambda}(\sigma_{1x})v_{\lambda}(\sigma_{2y})v_{\lambda}(\sigma_{3y})& =& -1 \hspace{0.7cm}\lambda \in F,\\
\label{eq5.63}
 v_{\lambda}(\sigma_{1y})v_{\lambda}(\sigma_{2x})v_{\lambda}(\sigma_{3y})& =& -1\hspace{0.7cm}\lambda \in G,\\
\label{eq5.64}
 v_{\lambda}(\sigma_{1y})v_{\lambda}(\sigma_{2y})v_{\lambda}(\sigma_{3x})& =& -1\hspace{0.7cm}\lambda \in H.
\end{eqnarray}
Now, since the only possible measurement outcomes for the spin operators are $+1$ or $-1$,
 the product of the quantities at the left hand side of the previous equations equals $+1$ while
 the product of the right hand sides is equal to $-1$.
Thus, Eqs.~(\ref{eq5.61})-(\ref{eq5.64}) are mutually inconsistent for any value of $\lambda$ belonging
 simultaneously to the subsets $E, F, G$, and $H$, that is, whenever $\lambda \in E \cap F \cap G\cap H$.
As a consequence, if a local deterministic hidden variable model exists and reproduces correctly
 the predictions of Eqs.~(\ref{eq5.61})-(\ref{eq5.64}), the subset $E \cap F \cap G\cap H$ must be
 the empty set. This implies either that anyone of the four subsets $E,F,G,$ or $H$, lies entirely in the
 complement of the intersection of the remaining three (e.g., $E \subseteq \Lambda - (F \cap G\cap H)$) or that
 the intersection of any two of the four subsets lies entirely in the complement of the intersection of
 the other two (e.g., $E \cap F \subseteq \Lambda - (G\cap H)$).
All these ten possibilities are easily proven to be completely equivalent from the point of view of the
 result we will obtain and, therefore, we will limit to consider only one of them, such as, for example,
 $E \subseteq \Lambda - (F \cap G\cap H)$. In this situation we have
\begin{eqnarray}
\label{eq5.7}
 \mu[ E] & \leq & \mu [ \Lambda- (F \cap G\cap H)] =\mu [ (\Lambda- F)\cup (\Lambda- G)\cup (\Lambda-H)]\\
 \label{eq5.72}
 & \leq & \mu [\Lambda - F]+\mu [\Lambda - G]+\mu [\Lambda - H]\\
 \label{eq5.73}
 & = & 3 -\mu[ F] -\mu [G] -\mu[H],
\end{eqnarray}
where the first inequality follows from that fact that $E \subseteq \Lambda - (F \cap G\cap H)$, while
 the other relations descend from trivial set theoretic properties.
Combining the above relation with Eqs.~(\ref{eq1.801})-(\ref{eq1.804}), one obtains the inequality
\begin{equation}
 \label{eq5.8} 4\varepsilon -1 \geq 0
\end{equation}
as a necessary condition for the existence of a local realistic model for any mixed state
 $\varrho$ having a trace distance $\varepsilon$ from the pure GHZ state and reproducing
 Eqs.~(\ref{eq1.41})-(\ref{eq1.44}).
Our derivation can be summarized in the following theorem: \\

{\bf Theorem}. Given the state vector $\vert GHZ \rangle = \frac{1}{\sqrt{2}}(\vert 0\rangle\vert 0\rangle
 \vert 0 \rangle + \vert 1\rangle \vert 1 \rangle\vert 1\rangle$ belonging to ${\mathbb{C}}^{2}\otimes {\mathbb{C}}^{2}
 \otimes {\mathbb{C}}^{2}$, let us consider a mixed statistical operators $\varrho$ whose trace distance
 $D(\varrho, \vert GHZ\rangle \langle GHZ \vert)$ from the pure GHZ state we denote with $\varepsilon$.
If there exists a local and deterministic hidden variable model for $\varrho$ then
 $ 4\varepsilon -1 \geq 0$. \\

The usefulness of this result is provided by the converse statement. In fact, the theorem implies
 that no local and deterministic hidden variable model can exist for all those (mixed)
 statistical operators whose trace distance $\varepsilon$ from the (pure) GHZ state is strictly less
 than $1/4$.
Equivalently, from a topological point of view, all mixed states lying in the interior of a ball of radius $1/4$
 centered in the (pure) GHZ state $\vert GHZ \rangle \langle GHZ\vert$ do not admit
 local realistic models.
Moreover, such states are not even completely separable, that is, they cannot be totally decomposed
 as a convex sum of single-particle product states (in fact, if they were completely separable, a local realistic
 model for them would exist).
To conclude, we have been able to generalize the GHZ argument to a whole class of tripartite mixed
 states, all exhibiting genuine nonlocal properties.


\section{The GHZ nonlocality argument in the presence of specific noises}

In order to point up nonlocal effects by means of an experimental implementations of the original GHZ
 argument~\cite{ghz}, one has to test the existence of
 perfect (anti)correlations (\ref{eq1.21})-(\ref{eq1.24}) between the outcomes of appropriately chosen single-particle
 observables~\cite{bouwmeester,pan}.
Unfortunately, in real experiments one will never observe the desired perfect match between
 measurement outcomes, for several reasons of a practical nature.
The first, which we are analyzing in this section, is due the mixedness of the
 quantum state under examination: imperfect state preparation procedures and unavoidable couplings
 with the environment produce in fact a (convex) mixture of states which is different from the pure
 GHZ state which is requested by the original argument of nonlocality.
Nonetheless, provided the resulting mixed state is close enough (in the trace topology)
 to the GHZ state, one can still run an experiment exhibiting nonlocal effects, as we have proved.
Another reason which defies the possibility of measuring perfect (anti)correlated outocomes derives
 from the fact that one (or more) observers may fail to perform a perfect measurement of the appropriate
 spin-observable.
Of course, spin-measurements performed along wrong directions result in non-perfect
 correlations between the measurement outcomes, and the issue of how to perform anyway a nonlocality test
 will be analyzed in the next section.

Let us start by considering the case of imperfect preparations affecting the GHZ state and
 take into account a particular kind of one-parameter class of mixed states, i.e.,
\begin{equation}
\label{eq6.1}
 \varrho = p | GHZ \rangle\langle GHZ | +
  \frac{1-p}{8}I_{2}\otimes I_{2}\otimes I_{2}
\end{equation}
obtained by making a convex sum of the pure GHZ state with the identity operator (also known as
 a white noise), the parameter $p\in [0,1]$ measuring the degree of purity of the state.
In spite of the fact that we are considering an impure GHZ state, the mixed state of Eq.~(\ref{eq6.1}) can
 still exhibit nonlocal features: in fact, due to the theorem we proved in the previous section, it is
 sufficient that the trace distance of $\varrho$ from the pure GHZ state is (strictly) less than $1/4$.
Since in this case $D(\varrho, \vert GHZ \rangle \langle GHZ \vert)\equiv \varepsilon = 7(1-p)/8$, there follows
that for any $p\in (5/7,1]$ nonlocal effects can still be exhibited by states of the form of
 Eq.~(\ref{eq6.1}), despite their lack of perfect (anti)correlations.

A larger interval of values of $p$ can be obtained if, instead of resorting to
 the evaluation of the trace distance as an upper bound for the difference between two
 probability distributions, we resort to direct calculation of the probabilities
 of Eqs.~(\ref{eq1.41})-(\ref{eq1.44}) and use the specific form of the state $\varrho$.
In fact, the knowledge of the precise form of the mixed state $\varrho$ allows us to calculate exactly
 the values of the relevant probability distributions, without resorting to
 the (not-always optimal) majorizations, like that of Eq.~(\ref{eq1.3}).
Thus, given the state of Eq.~(\ref{eq6.1}), we easily obtain the following relations:
\begin{eqnarray}
\label{eq6.21}
 P_{\varrho}(\sigma_{1x}\otimes \sigma_{2x} \otimes \sigma_{3x}=+1) & = & 1 -\eta,  \\
 \label{eq6.22}
P_{\varrho}(\sigma_{1x}\otimes \sigma_{2y} \otimes \sigma_{3y}=-1) & = & 1 -\eta, \\
\label{eq6.23}
P_{\varrho}(\sigma_{1y}\otimes \sigma_{2x} \otimes \sigma_{3y}=-1) & = & 1 - \eta, \\
\label{eq6.24}
 P_{\varrho}(\sigma_{1y}\otimes \sigma_{2y} \otimes \sigma_{3x}=-1) & = & 1 -\eta,
\end{eqnarray}
where $\eta =(1-p)/2$. Since $\eta$ is a smaller value than the trace distance between $\varrho$
 and the GHZ state, we expect to determine a larger interval of values of $p$ for which the associated
 statistical operators exhibit nonlocal properties.
In fact, if we proceed in exactly the same way as in the previous section, we conclude that the inequality
 $4\eta -1 \geq 0$ must be satisfied if a local and realistic hidden variable model exists for $\varrho$.
Therefore, the following particular result has been proved:\\

{\bf Theorem}. Given the state vector $\vert GHZ \rangle = \frac{1}{\sqrt{2}}(\vert 0\rangle\vert 0\rangle
 \vert 0 \rangle + \vert 1\rangle \vert 1 \rangle\vert 1\rangle$ belonging to ${\mathbb{C}}^{2}\otimes {\mathbb{C}}^{2}
  \otimes {\mathbb{C}}^{2}$, let us consider the mixed statistical operator $ \varrho = p | GHZ \rangle\langle GHZ | +
  \frac{1-p}{8}I_{2}\otimes I_{2}\otimes I_{2}$ where $p\in [0,1]$.
 If there exists a local and deterministic hidden variable model for $\varrho$ then
 $ p \in [0, \frac{1}{2}]$.\\

Equivalently, whenever $p\in (1/2,1]$ the corresponding statistical operator $\varrho$ does
 not admit any local realistic model and, as a consequence, genuine nonlocal effects
 can be revealed by experiments aiming to verify the probability distributions
 of Eqs.~(\ref{eq6.21})-(\ref{eq6.24}). Once again, even in this situation, the presence of a completely
 chaotic noise has not necessarily destroyed the possibility of setting up an experimental
 test of a GHZ-like argument.

An equivalent result can be obtained by replacing the completely chaotic noise with a
 colored noise.
This kind of noise has been recently suggested~\cite{cabello} to be the best choice for describing
 entangled states produced in type II spontaneous parametric down conversion experiments~\cite{kwiat}.
Therefore, a more realistic (with respect to the white one which is usually considered
 in the literature~\cite{white}) kind of noise affecting the
 preparation of a pure GHZ state can be represented by the following class of one-parameter
 statistical operators $\varrho$:
\begin{equation}
\label{eq6.3}
 \varrho= p | GHZ \rangle\langle GHZ | + \frac{1-p}{2}(\vert 000 \rangle \langle 000 \vert +
 \vert 111 \rangle \langle 111\vert).
\end{equation}
A simple calculation shows that the probabilities of Eqs.~(\ref{eq6.21})-(\ref{eq6.24})
 remain unaltered and, as a consequence, the same conclusions we obtained in presence of a white noise
 hold true for the case of a colored noise as well.\\

{\bf Theorem}. Given the state vector $\vert GHZ \rangle = \frac{1}{\sqrt{2}}(\vert 0\rangle\vert 0\rangle
 \vert 0 \rangle + \vert 1\rangle \vert 1 \rangle\vert 1\rangle$ belonging to ${\mathbb{C}}^{2}\otimes {\mathbb{C}}^{2}
 \otimes {\mathbb{C}}^{2}$, let us consider the mixed statistical operator $ \varrho = p | GHZ \rangle\langle GHZ |
 +\frac{1-p}{2}(\vert 000 \rangle \langle 000 \vert + \vert 111 \rangle \langle 111\vert)$ where $p\in [0,1]$.
If there exists a local and deterministic hidden variable model for $\varrho$ then $p \in [0, \frac{1}{2}]$.


\section{Imperfect measurements}

As remarked in the previous section, even if a pure GHZ state is considered,
 perfect (anti)correlated outcomes are not obtained whenever the observers perform imperfect
 measurements of the spin-observables they are requested to measure.
In fact, the original GHZ argument~\cite{ghz} requires the three space-like separated
 observers to measure the spin of their particle along the $x$ and $y$ axes, defined with
 respect to a common reference frame.
Imperfect measurements therefore may arise when the measurements are performed along axes
 which are different from those prescribed by the experiment and, as a consequence, (anti)correlated
 outcomes are no more guaranteed to occur.
To start with, let us suppose that, while the state is the pure GHZ state, observer $1$, for example,
 performs imperfect measurements, measuring the spin along the (wrong and completely arbitrary)
 directions ${\bar{x}}$ and ${\bar{y}}$ instead of $x$ and $y$.
Given an orthogonal three-dimensional coordinate frame, we may suppose, without
 loss of generality, that the unit vector $\bar{x}$ lies in the $xz$ plane while $\bar{y}$
 may be directed arbitrarily in space, so that
\begin{eqnarray}
\label{eq7.1}
\sigma_{\bar{x}} & = & \cos{\theta} \:\sigma_{x} +\sin{\theta}\: \sigma_{y}, \\
\sigma_{\bar{y}} & = & \sin{\beta}\cos{\alpha} \:\sigma_{x} +\sin{\beta} \sin{\alpha} \:\sigma_{y}
+\cos{\beta}\:\sigma_{z},
\end{eqnarray}
where $\theta,\alpha\in [0,2\pi)$ while $\beta \in [0,\pi]$ are the angles individuating the
 directions $\bar{x}$ and $\bar{y}$ with respect to the reference frame. Given the explicit
form of the new spin-observables, cumbersome but straightforward calculations
 tell us how to modify the probability distribution of Eqs.~(\ref{eq1.21})-(\ref{eq1.24}) in case
 of imperfect measurements performed by the first observer,
\begin{eqnarray}
\label{eq7.21}
 P_{GHZ}(\sigma_{1\bar{x}}\otimes \sigma_{2x} \otimes \sigma_{3x}=+1) & = & 1- (1-\cos{\theta})/2, \\
 \label{eq7.22}
P_{GHZ}(\sigma_{1\bar{x}}\otimes \sigma_{2y} \otimes \sigma_{3y}=-1) & = & 1- (1-\cos{\theta})/2,\\
\label{eq7.23}
P_{GHZ}(\sigma_{1\bar{y}}\otimes \sigma_{2x} \otimes \sigma_{3y}=-1) & = & 1- (1-\sin{\alpha}\sin{\beta})/2,\\
\label{eq7.24}
 P_{GHZ}(\sigma_{1\bar{y}}\otimes \sigma_{2y} \otimes \sigma_{3x}=-1) & = & 1-(1-\sin{\alpha}\sin{\beta})/2.
\end{eqnarray}
Let us proceed as usual and suppose that a local realistic description exists for the GHZ state
 reproducing the previous equations, and define the subsets $E,F,G$, and $H$ of $\Lambda$ as in
 Eq.~(\ref{eq1.6}) and~(\ref{eq1.61})-(\ref{eq1.63}).
The probability distributions of Eqs.~(\ref{eq7.21})-(\ref{eq7.24}) require the following values for
 the measure of the $E,F,G,H$ subsets of $\Lambda$:
\begin{eqnarray}
\label{eq7.31}
 \mu[ E ] = \mu [F]  & = & 1- (1-\cos{\theta})/2, \\
 \label{eq7.32}
\mu[ G] = \mu [H] & = & 1- (1-\sin{\alpha}\sin{\beta})/2.
\end{eqnarray}
Equation~(\ref{eq5.73}), together with the new values of Eqs.~(\ref{eq7.31}-\ref{eq7.32}) for the measure
 of the considered subsets, implies the following constraint
 between the angles $\theta,\alpha$ and $\beta$ (identifying the imperfect measurement directions)
 when a local realistic description for the GHZ state exists:
\begin{equation}
\label{eq7.4} \cos{\theta} +\sin{\alpha}\sin{\beta} \leq 1.
\end{equation}
As a consequence, whenever one of the observers makes imperfect measurements of
 his spin-observables, is it still possible to deny the existence of a local realistic
 model for the (pure) GHZ state as far as $\cos{\theta} +\sin{\alpha}\sin{\beta} > 1$.
Of course, the relation~(\ref{eq7.4}) turns out to be maximally violated when
 $\cos{\theta}=\sin{\alpha}\sin{\beta}=1$, that is when $\bar{x}=x$ and $\bar{y}=y$, as
 expected.

Thus, imperfect measurements do not necessarily defy the possibility of putting into evidence
 the nonlocal effects of the GHZ state. Proof of this has been obtained by applying
 the same arguments we exhibited in Section III where we were dealing with the case of mixed states.
Moreover nonlocality can also be proven, by resorting to the same logical arguments but
 with more cumbersome calculations, when we suppose that more than one observer performs
 imperfect measurements or when considering the simultaneous effect of imperfect state
 preparations and imperfect measurements.


\section{Conclusions}

In this paper we have exhibited a set-theoretic generalization of the GHZ proof of nonlocality
 without inequalities which covers the case of mixed statistical operators.
More precisely, a necessary condition for the existence of a local and deterministic hidden variable model
 reproducing the quantum mechanical predictions of any tripartite spin-$1/2$ mixed state
 has been determined.
Failure of this condition implies the impossibility of obtaining a local realistic description
 for the associated statistical operators and, as a consequence, that such operators cannot be
 decomposed in a convex sum of single-particle product states.
Finally, as a practical application of our condition, we have determined the maximum amount
 of (appropriately chosen kind of) noise, affecting the experimental realization of the
 original GHZ argument, such that it is still possible to highlight genuine nonlocal effects.
The analogous problem arising as a consequence of the unavoidable imperfect measurement procedures
 has been also analyzed.

\section{Acknowledgments}

This work was supported in part by Istituto Nazionale di Fisica Nucleare, Sezione di Trieste, Italy.


\end{document}